# Computational study of flow dynamics from a dc arc plasma jet

**Juan Pablo Trelles**

Department of Mechanical Engineering, University of Massachusetts Lowell, Lowell, MA 01854, USA

E-mail: Juan_Trelles@uml.edu

**Abstract**

Plasma jets produced by direct-current (DC) non-transferred arc plasma torches, at the core of technologies ranging from spray coating to pyrolysis, present intricate dynamics due to the coupled interaction of fluid flow, thermal, and electromagnetic phenomena. The flow dynamics from an arc discharge plasma jet are investigated using time-dependent three-dimensional simulations encompassing the dynamics of the arc inside the torch, the evolution of the jet through the discharge environment, and the subsequent impingement of the jet over a flat substrate. The plasma is described by a chemical equilibrium and thermodynamic nonequilibrium (two-temperature) model. The numerical formulation of the physical model is based on a monolithic and fully-coupled treatment of the fluid and electromagnetic equations using a Variational Multiscale Finite Element Method. Simulation results uncover distinct aspects of the flow dynamics, including the jet forcing due to the movement of the electric arc, the prevalence of deviations between heavy-species and electron temperatures in the plasma fringes, the development of shear flow instabilities around the jet, the occurrence of localized regions with high electric fields far from the arc, and the formation and evolution of coherent flow structures.

## 1. Introduction

*1.1. Flow dynamics from arc plasma torches*

Direct-current (DC) non-transferred arc plasma torches are at the core of diverse technologies, such as plasma spray coating, chemical and powder synthesis, extractive metallurgy, toxic waste treatment, and pyrolysis [1, 2, 3]. Plasma jets in these technologies are used as directed sources of very high energy, momentum, and excited species fluxes to a target material or workpiece. The study of the interaction of a thermal plasma jet with a substrate is particularly relevant when maximum utilization of the fluxes from the plasma is sought, e.g., to process large volumes of material, as required in waste treatment and pyrolysis.

The plasma jet from a DC non-transferred arc plasma torch is formed by the interaction of a stream of working fluid (typically an inert or molecular gas) with an electric arc established between a (semi-) conical cathode and a cylindrical anode surrounding it [2]. Figure 1 schematically depicts the inside of the torch, the interaction of the arc with the inflow of working gas, and the ejection of the plasma to the environment outside the torch forming the jet.

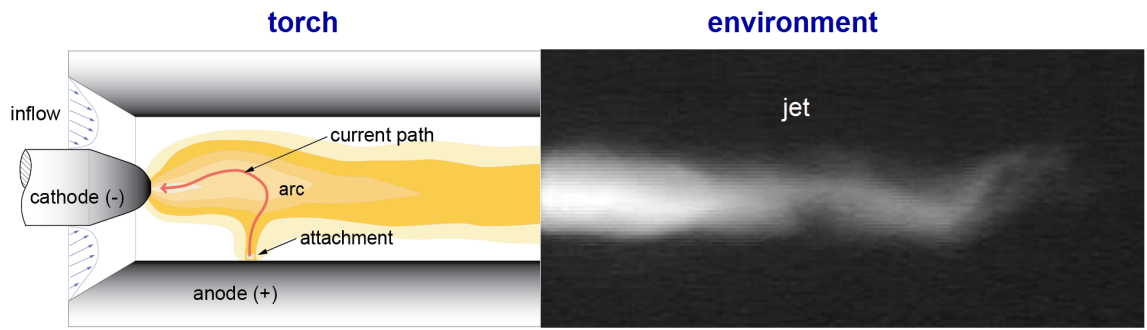

**Figure 1.** Flow dynamics from an arc plasma jet: (*left*) schematic of the plasma inside a DC non-transferred arc torch and (*right*) optical high-speed image of the plasma jet.

Diverse physical phenomena drive the flow dynamics of the jets from non-transferred arc torches. The forcing of the jet produced by the quasi-periodic movement of the arc inside the torch is a primary factor affecting the flow dynamics, as partially evidenced by undulant shape of the jet in Fig. 1 and experimentally measured voltage signals [4]. The arc dynamics are a result of the unstable imbalance between electromagnetic (Lorentz) and flow drag forces found under most industrially relevant operating conditions (i.e., relatively large ratios of flow rate over total electric current). In addition, the presence of large field and material property gradients, typical of the interaction of arc discharges with a surrounding media, lead to various types of fluid dynamic, thermal, and electromagnetic instabilities [4, 5]. These instabilities affect the arc dynamics (e.g., kink and sausage electromagnetic instabilities) as well as the evolution of the jet (e.g., shear-flow instabilities in the periphery of the jet), and drive the large- and small-scale evolution of the plasma flow. Therefore, the adequate analysis of flow dynamics from DC arc jets requires the coupled description of not only the jet, but also of the arc inside the torch.

The study of flow dynamics from DC arc plasma jets, compared to other thermal plasma flows, is particularly involved due to the incidence of thermodynamic nonequilibrium effects and the development of flow turbulence [2, 6-8]. Thermodynamic nonequilibrium (i.e., dissimilar relaxation of heavy-species and electron kinetic energy) is found when the plasma interacts with its surroundings, such as a cold gas environment or a solid boundary; and its extent increases with the strength of the interaction (e.g., higher degree of nonequilibrium is often found for higher ratios of flow rate over total current). Flow turbulence is often a direct consequence of the evolution of fluid instabilities. Of particular relevance in DC arc plasma jets is the occurrence of shear-flow (Kelvin-Helmholtz) instabilities, which are seeded at the interface between parallel streams of gas with markedly dissimilar velocity and/or material properties (e.g., density, viscosity), as found at the boundary of the plasma jet with the discharge environment.

*1.2. Computational modeling of plasma flows*

Numerical modeling and simulation provide unique capabilities for the detailed analysis of plasma flow dynamics [9]. There are several reports of the use of models based on the Local Thermodynamic Equilibrium (LTE) assumption to describe the dynamics inside DC arc plasma torches (e.g., [10-16]), and few that describe the arc dynamics together with the plasma jet [17, 18]. The strong interaction of the plasma with the discharge environment makes the use of models that abandon the LTE assumption in favor of a non-LTE (NLTE) description particularly suitable for the description of DC arc plasma jets. Nevertheless, the increased complexity and computational cost of NLTE models compared to LTE ones have limited the widespread use of the former compared to the later. Notable exceptions are the work reported in [19] of the arc dynamics inside plasma torches, and the work by Colombo and collaborators [20] of the LTE and NLTE modeling of a twin plasma torch. Preliminary computational studies of the flow dynamics from DC arc plasma jets using a NLTE model were reported in [21, 22].

The accurate description of flow turbulence adds a significantly higher degree of complexity to plasma flow modeling. There are numerous reports of the use of Reynolds-Averaged Navier-Stokes turbulence models for the analysis of thermal plasma jets, such as the work of Huang *et al* [23] on the two-dimensional simulation of a jet using a two-fluid model to describe the entrainment of the cold gas from the surrounding environment into the plasma jet, or the work of Li and Chen [24] on the three-dimensional simulation of the impingement of a plasma jet using a k-ε turbulent model. Time-dependent three-dimensional models are more suitable to describe the intrinsically dynamic and three-dimensional flow from arc plasma jets. Colombo *et al* [25] performed Large Eddy Simulations (LESs), arguably the state-of-the-art coarse-grained turbulence modeling approach, to the simulation of the flow from an inductively coupled plasma torch into a reaction chamber. More recently, Shigeta [26] presented a computational study of the flow from a radio-frequency inductively coupled plasma interacting with a DC plasma jet. Shigeta's simulations did not make explicit use of any turbulence model, which could arguably be considered as so-called Direct Numerical Simulations (DNSs). (A DNS attempts to describe all the scales of a turbulent flow, and therefore do not require any type of turbulence model.) The simulations in [26] captured the development of coherent flow structures, which could otherwise not be captured if the model presents significant or inadequate numerical dissipation (e.g., due to insufficient spatial resolution or the use of low accuracy numerical discretization schemes). The time-dependent three-dimensional NLTE simulations in [21, 22] did not include any turbulence model, yet the limited spatial and temporal resolution used prevents them to be considered DNS.

*1.3. Scope and organization of the study*

This article reports the study of flow dynamics from a DC arc plasma jet using a chemical equilibrium and thermodynamics nonequilibrium (NLTE or two-temperature) plasma model. The NLTE model explicitly describes the evolution and interaction of the volumetric energy density associated with the heavy-species (i.e., atoms and ions) and that associated with electrons by using different energy conservation equations for the heavy-species and electrons, respectively. The model is applied to the three-dimensional and time-dependent description the flow dynamics from an argon plasma discharging in an argon environment. The simulations encompass the dynamics of the arc inside the torch, the evolution of the jet through the discharge environment, and the subsequent impingement of the jet over a flat substrate.

The numerical formulation of the physical model is based on a monolithic and fully-coupled treatment of the fluid and electromagnetic equations using a Variational Multiscale Finite Element Method (VMS-FEM) [27-30]. Even though the VMS method has been applied for the complete and consistent LES of other types of flows (e.g., [28]), no attempt to model turbulent effects has been pursued in the present study. The discrete problem is solved numerically using a generalized-alpha predictor multi-corrector time stepper [31] together with an inexact Newton-Krylov with line search nonlinear solver [32-34]. A comprehensive description of the numerical model used and its validation is found in [35].

The simulation results capture different aspects of the flow dynamics, including the jet forcing caused the dynamics behavior of the arc, the predominance of thermodynamic nonequilibrium in the plasma fringes, the development of shear flow instabilities around the jet, the occurrence of localized regions with high electric fields far from the arc, the fluctuating expansion of the gas ejected from the torch, and the formation and evolution of coherent flow structures resembling the experimental findings reported in [6].

The paper is organized as follows: Section 2 presents the mathematical plasma flow model based on a fully-coupled monolithic treatment of the thermodynamic nonequilibrium and chemical equilibrium fluid model together with the electromagnetic field evolution equations. Section 3 describes the numerical model based on the VMS-FEM and the numerical solution approach. Section 4 describes the simulation set-up: the geometry of the spatial domain, the boundary conditions used, and the implementation of nonreflecting conditions to mitigate wave reflection from outflow boundaries. Section 5 presents the computational results of the simulation of the plasma flow from a DC arc torch. The summary and conclusions of the study are presented in Section 6.

## 2. Mathematical model

*2.1. Flow evolution equations*

The mathematical plasma flow model is based on a fully-coupled treatment of a fluid flow model together with the electromagnetic field evolution equations. The assumptions used for the model are:

- A fluid model, in contrast to a particle-based model, is adequate given the relatively high collision frequencies among the constituent particles in high-pressure (e.g., atmospheric) arc discharge plasmas.
- The plasma is described as a compressible, reactive, electromagnetic fluid.
- The constituent species in the plasma are assumed in chemical equilibrium.
- The electrons and heavy-species obey different Maxwellian velocity distributions, and therefore the plasma is considered in a state of thermodynamic nonequilibrium (NLTE).
- The plasma is considered non-relativistic, non-magnetized, quasi-neutral, and the macroscopic Maxwell's equations provide a suitable description of the evolution of electromagnetic fields.
- Charge transport is dominated by the electric field distribution and by electron diffusion; ion diffusion and Hall currents are assumed negligible.
- Radiative transport is modeled assuming the plasma is optically thin.

Based on the above assumptions, the plasma flow is described by the set of equations for: (1) conservation of total mass, (2) conservation of mass-averaged linear momentum, (3) conservation of thermal energy of heavy-species, (4) conservation of thermal energy of electrons, (5) conservation of electric charge, and (6) magnetic induction. These equations are summarized in Table 1 as a single set of *transient – advective – diffusive – reactive* (TADR) transport equations.

**Table 1.** Set of fluid – electromagnetic evolution equations for the arc discharge plasma flow; for each equation: *Transient + Advective – Diffusive – Reactive* = 0.

| Equation | Transient | Advective | Diffusive | Reactive |
|---|---|---|---|---|
| Conservation of total mass | $\partial_t \rho$ | $\mathbf{u} \cdot \nabla \rho + \rho \nabla \cdot \mathbf{u}$ | 0 | 0 |
| Conservation of linear momentum | $\rho \partial_t \mathbf{u}$ | $\rho \mathbf{u} \cdot \nabla \mathbf{u} + \nabla p$ | $\nabla \cdot \mu (\nabla \mathbf{u} + \nabla \mathbf{u}^T) - \nabla \cdot (\frac{2}{3}\mu (\nabla \cdot \mathbf{u}) \boldsymbol{\delta})$ | $\mathbf{J}_q \times \mathbf{B}$ |
| Thermal energy heavy-species | $\rho \partial_t h_h$ | $\rho \mathbf{u} \cdot \nabla h_h$ | $\nabla \cdot (\kappa_{hr} \nabla T_h)$ | $\partial_t p_h + \mathbf{u} \cdot \nabla p_h + K_{eh}(T_e - T_h)$ |
| Thermal energy electrons | $\rho \partial_t h_e$ | $\rho \mathbf{u} \cdot \nabla h_e$ | $\nabla \cdot (\kappa_e \nabla T_e)$ | $\partial_t p_e + \mathbf{u} \cdot \nabla p_e - K_{eh}(T_e - T_h) - 4\pi \varepsilon_r + \mathbf{J}_q \cdot (\mathbf{E} + \mathbf{u} \times \mathbf{B}) + \frac{5 k_B}{2e} \mathbf{J}_q \cdot \nabla T_e$ |
| Conservation of electric charge | 0 | 0 | $\nabla \cdot (\sigma \nabla \phi_p) - \nabla \cdot (\sigma \mathbf{u} \times (\nabla \times \mathbf{A}))$ | 0 |
| Magnetic induction | $\mu_0 \sigma \partial_t \mathbf{A}$ | $\mu_0 \sigma \nabla \phi_p - \mu_0 \sigma \mathbf{u} \times (\nabla \times \mathbf{A})$ | $\nabla^2 \mathbf{A}$ | 0 |

In Table 1, $\partial_t \equiv \partial / \partial t$ is the partial derivative with respect to time, $\nabla$ and $\nabla \cdot$ are the gradient and the divergence operators, respectively; $\rho$ represents total mass density, $\mathbf{u}$ mass-averaged velocity, $p$ total pressure, $\mu$ is the dynamic viscosity, the superscript $^T$ indicates the transpose operator, and $\boldsymbol{\delta}$ the Kronecker delta tensor. The conservation equations for momentum and thermal energy in Table 1 are written in the so-called advective form as they implicitly imply total mass conservation (i.e., $\partial \rho + \nabla \cdot (\rho \mathbf{u}) = 0$).

The diffusive term in the conservation of linear momentum equation describes the divergence of the stress tensor $\boldsymbol{\tau}$, therefore:

$$\boldsymbol{\tau} = -\mu(\nabla \mathbf{u} + \nabla \mathbf{u}^\top - \tfrac{2}{3}\mu(\nabla \cdot \mathbf{u})\boldsymbol{\delta}), \tag{1}$$

where the $\tfrac{2}{3}$ factor next to the fluid dilatation term $\nabla \cdot \mathbf{u}$ arises from the use of Stoke's hypothesis for the dilatational viscosity [35]. The reactive term in the momentum conservation equation represents the Lorentz force, where $\mathbf{J}_q$ is the electric current density and $\mathbf{B}$ the magnetic field.

The thermal energy conservation equations are written in terms of $h_h$ and $h_e$, which stand for the enthalpy of the heavy-species and electrons, respectively. The diffusive term in the conservation of thermal energy of heavy-species represents the divergence of the total heavy-species energy transported by diffusion, therefore:

$$-\kappa_{hr}\nabla T_h = -\kappa_h \nabla T_h + \sum_{s \neq e} h_s \mathbf{J}_s, \tag{2}$$

where $T_h$ is the heavy-species (translational) temperature, $\kappa_{hr}$ represents the *translational-reactive* heavy-species thermal conductivity, $\kappa_h$ is the *translational* heavy-species conductivity, $\mathbf{J}_s$ and $h_s$ stand for the diffusive mass transport flux and specific enthalpy of species $s$, respectively; and the summation in Eq. 2 runs over all the heavy-species in the plasma (i.e., all species except electrons). The first term in right-hand-side of Eq. 2 represents the heat transported by conduction, whereas the second term the enthalpy transported by mass diffusion and if often modeled by defining a *reactive* thermal conductivity $\kappa_r$ such that $\sum_{s \neq e} h_s \mathbf{J}_s \approx -\kappa_r \nabla T_h$. Given the chemical equilibrium assumption, the coefficients $\kappa_{hr}$ and $\kappa_r$ can be treated as any other transport properties (e.g., see [37]).

The diffusive term in the electron thermal energy conservation equation describes the transport of electrons by conduction, where $T_e$ is the electron temperature and $\kappa_e$ the electron translational thermal conductivity. The terms $\partial_t p_h + \mathbf{u} \cdot \nabla p_h$ and $\partial_t p_e + \mathbf{u} \cdot \nabla p_e$ describe the heavy-species and electron pressure work, respectively, where $p_h$ and $p_e$ are the heavy-species and electron pressure, respectively. The term $K_{eh}(T_e - T_h)$ models the relaxation of heavy-species and electron energy, where $K_{eh}$ is the electron – heavy-species energy exchange coefficient (inversely proportional to a characteristic time for inter-particle energy exchange). The viscous heating term $-\boldsymbol{\tau} : \nabla \mathbf{u}$ in the heavy-species thermal energy conservation equation has been omitted because it is negligible in the plasma flow of interest. The term $4\pi\varepsilon_r$ in the electron energy conservation equation models the radiation losses from the plasma using the *effective net emission* approximation, where $\varepsilon_r$ is the effective net emission coefficient for an optically thin plasma. The term $\mathbf{J}_q \cdot (\mathbf{E} + \mathbf{u} \times \mathbf{B})$ represents Joule heating, where $\mathbf{E}$ is the electric field.

The last component of the reactive term in the electron thermal energy conservation equation arises from the transport of energy by electron diffusion. Considering that the electric current density is dominated by the transport of charge by electrons (due to their smaller mass and higher mobility), the energy transported by electron diffusion can be approximated by:

$$\mathbf{J}_e \approx -\frac{m_e}{e}\mathbf{J}_q, \qquad (3)$$

where $m_e$ is the electron mass and $e$ represents the elementary charge. Therefore,

$$\nabla \cdot (h_e \mathbf{J}_e) \approx -\frac{m_e}{e}(\mathbf{J}_q \cdot \nabla h_e + h_e \nabla \cdot \mathbf{J}_q) = -\frac{5 k_B}{2e} \mathbf{J}_q \cdot \nabla T_e, \qquad (4)$$

where charge conservation, $\nabla \cdot \mathbf{J}_q = 0$, has been invoked, as well as the definition of electron enthalpy $h_e = \tfrac{5}{2} k_B m_e T_e$, where $k_B$ is Boltzmann's constant.

The diffusive term in the electric charge conservation equation represents the divergence of the current density $\mathbf{J}_q$; hence:

$$\mathbf{J}_q = \sigma(\mathbf{E}_p + \mathbf{u} \times \mathbf{B}), \qquad (5)$$

where $\sigma$ is the electric conductivity and $\mathbf{E}_p$ represents the effective electric field given by:

$$\mathbf{E}_p = -\nabla \phi_p - \partial_t \mathbf{A}, \qquad (6)$$

where $\phi_p$ is the effective electric potential and $\mathbf{A}$ the magnetic vector potential defined from:

$$\nabla \times \mathbf{A} = \mathbf{B}. \qquad (7)$$

The use of $\phi_p$ and $\mathbf{A}$ allows the *a priori* satisfaction of the solenoidal constraint $\nabla \cdot \mathbf{B} = 0$ in Maxwell's equations. The Coulomb gauge condition $\nabla \cdot \mathbf{A} = 0$ is used to define $\mathbf{A}$ uniquely. The charge conservation equation assumes $\nabla \cdot (\sigma \partial_t \mathbf{A}) \approx 0$, as this term is negligible in the flow of interest and its omission greatly simplifies the implementation of the model by avoiding mixed spatial-temporal derivatives.

The *effective* electric field $\mathbf{E}_p$ is used to describe *generalized* Ohm's laws, which account for Hall effects and provide a more detailed description of charge transport due to diffusion processes. Consistent with the assumptions listed above, $\mathbf{E}_p$ and $\mathbf{E}$ are related by:

$$\mathbf{E}_p \approx \mathbf{E} + \frac{\nabla p_e}{e n_e}, \qquad (8)$$

where $p_e$ is the electron pressure and $n_e$ the number density of electrons. (Terms accounting for the transport of charge by ion diffusion and Hall effects have been neglected in Eq. 8.) It can be noticed that the Joule heating term involves **E** and not **E**$_p$. It is customary to assume **E**$_p \approx$ **E** in LTE models (and therefore neglect the effect of the second term in the right side of Eq. 8).

In the magnetic induction equation, $\mu_0$ represents the permeability of free space. Whereas the charge conservation equation in Table 1 restates Gauss' law, the magnetic induction equation combines Faraday's and Ampere's laws. Given the satisfaction of the solenoidal constraint implied by the use of the magnetic vector potential in Eq. 7, the charge conservation and magnetic induction equations enclose a complete description of Maxwell's equations.

*2.2. Monolithic plasma flow model*

The complete system of equations in Table 1 is treated in a fully-coupled monolithic manner as a single TADR transport system. This system is expressed in *residual* form as:

$$\mathcal{R}(\mathbf{Y}) = \underbrace{\mathbf{A_0}\partial_t \mathbf{Y}}_{\text{transient}} + \underbrace{(\mathbf{A}_i \partial_i)\mathbf{Y}}_{\text{advective}} - \underbrace{\partial_i(\mathbf{K}_{ij}\partial_j \mathbf{Y})}_{\text{diffusive}} - \underbrace{(\mathbf{S_1 Y} - \mathbf{S_0})}_{\text{reactive}} = \mathbf{0}, \qquad (9)$$

where $\mathcal{R}$ represents the residual of the system of equations, **Y** is the vector of unknowns, the sub-indexes *i* and *j* stand for each spatial coordinate (e.g., for three-dimensional Cartesian coordinates, the vector of spatial coordinates $\mathbf{X} = [x\ y\ z]^T$, and therefore $i, j = \{\ x, y, z\ \}$), and the Einstein summation convention of repeated indexes is used (e.g., the divergence of a field $\mathbf{a} = \{a_i\}$ is given by: $\nabla \cdot \mathbf{a} = \partial_x a_x + \partial_y a_y + \partial_z a_z \equiv \partial_i a_i$). The matrices $\mathbf{A_0}$, $\mathbf{A}_i$, $\mathbf{K}_{ij}$ and $\mathbf{S_1}$ are denoted as the *transient*, *advective*, *diffusive*, and *reactive* (TADR) transport matrices, respectively; which, given the non-linear nature of the model equations, are functions of **Y**.

Any independent set of variables can be chosen as components of the vector **Y** to seek solution of Eq. 9. In the present study, the vector **Y** is chosen as the set of *primitive* variables, i.e.,

$$\mathbf{Y} = [\ p\quad \mathbf{u}^T\quad T_h\quad T_e\quad \phi_p\quad \mathbf{A}^T\ ]^T. \qquad (10)$$

The set of variables composing the vector **Y** are mapped one-to-one to the set of equations in Table 1; i.e., the equation for conservation of total mass is used as the primary equation to find *p*, the momentum conservation equation for finding **u**, the heavy-species conservation equation for determining $T_h$, etc.

The set of variables in Eq. 10 is robust for the description of both, incompressible and compressible flows (e.g., $\rho$ could be used as independent variable instead of *p*, but its behavior is not well defined in the incompressible flow limit [38]). Furthermore, this set of variables allows the greatest solution accuracy for the variables of interest (e.g., the solution procedure aims to attain convergence of the heavy-species energy conservation equation directly in terms of $T_h$, which is the main variable to describe heavy-species energy; an alternative procedure could solve for $h_h$ and then find $T_h$ in an intermediate or post-processing step).

Once the vector $\mathbf{Y}$ is defined, the transport matrices can be expressed completely in terms of $\mathbf{Y}$. For example, the transient term in conservation equation for total mass can be expressed in terms of the variables in Eq. 10 as:

$$\partial_t \rho = (\frac{\partial \rho}{\partial p})\partial_t p + (\frac{\partial \rho}{\partial T_h})\partial_t T_h + (\frac{\partial \rho}{\partial T_e})\partial_t T_e, \tag{11}$$

where the first term in the right-hand-side describes acoustic propagation (e.g., negligible or ill defined in incompressible flows), and the second and third terms, the dependence of mass density in heavy-species and electron temperatures, responsible for heat wave expansion. Explicit expressions for the matrices $\mathbf{A_0}$, $\mathbf{A_i}$, $\mathbf{K}_{ij}$ and $\mathbf{S_1}$ and the vector $\mathbf{S_0}$ composing the plasma flow model are presented in [35].

Given the set of plasma flow equations in Table 1 and the set of independent variable in Eq. 10, closure of the physical - mathematical model requires the definition of thermodynamic ($\rho$, $\partial \rho / \partial p$, $\partial \rho / \partial T_h$, $\partial \rho / \partial T_e$, $h_h$, $\partial h_h / \partial p$, $\partial h_h / \partial T_h$, etc.) and transport ($\mu$, $\kappa_{hr}$, $\kappa_e$, $\sigma$) material properties, as well as of the terms $K_{eh}$ and $\varepsilon_r$.

*2.3. Material properties and constitutive relations*

The calculation of material properties for a plasma in chemical equilibrium is composed of three consecutive steps: (1) calculation of the plasma composition, (2) calculation of thermodynamics properties, and (3) calculation of transport properties.

The plasma composition is determined by using the mass action law (minimization of Gibbs free energy), Dalton's law of partial pressures, and the quasi-neutrality condition [1]. The present study considers a four-species argon plasma, composed of the species: $Ar$, $Ar^+$, $Ar^{++}$, and $e^-$; therefore, the equations used to determine the number density $n_s$ of each species $s$ are:

$$\frac{n_{e^-} n_{Ar^+}}{n_{Ar}} = \frac{Q_{e^-} Q_{Ar^+}}{Q_{Ar}} (\frac{2\pi m_e T_e}{h_P^2})^{\frac{3}{2}} \exp(-\frac{\varepsilon_{Ar^+}}{k_B T_e}), \tag{12}$$

$$\frac{n_{e^-} n_{Ar^{++}}}{n_{Ar^+}} = \frac{Q_{e^-} Q_{Ar^{++}}}{Q_{Ar^+}} (\frac{2\pi m_e T_e}{h_P^2})^{\frac{3}{2}} \exp(-\frac{\varepsilon_{Ar^{++}}}{k_B T_e}), \tag{13}$$

$$n_{Ar} + n_{Ar^+} + n_{Ar^{++}} + \theta n_{e^-} = \frac{p}{k_B T_h}, \text{ and} \tag{14}$$

$$n_{Ar^+} + n_{Ar^{++}} - n_{e^-} = 0; \tag{15}$$

where $h_P$ is Planck's constant, $Q_s$ and $\varepsilon_s$ are the partition function and formation (ionization) energy of species $s$, respectively [1, 39]; and $\theta = T_e/T_h$ is the so-called thermodynamic nonequilibrium parameter. Equations 14 and 15 are Saha equations appropriate for the NLTE model in which the lowering of the ionization energy has been neglected.

Once the plasma composition is known, the thermodynamic properties are calculated by:

$$\rho = \sum_s m_s n_s, \qquad (16)$$

$$p_h = k_B (n_{Ar} + n_{Ar^+} + n_{Ar^{++}}) T_h, \qquad (17)$$

$$p_e = k_B n_{e^-} T_e, \qquad (18)$$

$$h_h = \rho^{-1} \sum_{s \neq e} (\tfrac{5}{2} k_B n_s T_h + n_s \varepsilon_s + k_B n_s T_e \frac{dQ_s}{d \ln T_e}), \text{ and} \qquad (19)$$

$$h_e = \rho^{-1} \tfrac{5}{2} k_B n_{e^-} T_e, \qquad (20)$$

where $m_s$ is the mass of species $s$.

The accurate calculation of thermodynamic nonequilibrium transport properties for a plasma following the Chapman-Enskog procedure [40] is computationally demanding, particularly within time-dependent three-dimensional simulations. To reduce the computational cost, the NLTE model for the 4-component argon plasma uses look-up tables based on the nonequilibrium transport properties at $p = 1$ [atm] reported in [41, 42].

The net emission coefficient $\varepsilon_r$ is calculated as a function of $T_e$ using the values reported in [44] for an optically thin argon plasma within a table look-up procedure. The volumetric electron – heavy species energy exchange coefficient $K_{eh}$ is modeled as:

$$K_{eh} = \sum_{s \neq e} \tfrac{3}{2} k_B \frac{2 m_e m_s}{(m_s + m_e)} \left(\frac{8 k_B T_e}{\pi m_e}\right)^{\frac{1}{2}} n_s \sigma_{es}, \qquad (21)$$

where $\sigma_{es}$ is the collision cross-section between electrons and the heavy-species $s$, which is calculated using the Coulomb collision cross-section for electron – ion collisions and the data in [43] for the electron – neutral collision. A detailed description of the calculation of material properties and constitutive relations is presented in [44].

## 3. Numerical model

*3.1. Variational multiscale finite element method*

The mathematical plasma flow model given Eq. 9, complemented with appropriate initial and boundary conditions specified over the spatial domain $\Omega$ with boundary $\Gamma$, can be cast in the so-called *weak*, or variational, as:

$$\int_\Omega \mathbf{W}\,\mathcal{R}(\mathbf{Y})d\Omega = (\mathbf{W},\mathcal{R}(\mathbf{Y}))_\Omega = \mathbf{0}, \tag{22}$$

where $\mathbf{W}$ is any function that belongs to the same mathematical space of $\mathbf{Y}$.

The direct numerical solution of Eq. 9 or Eq. 22 is posed with diverse spurious behavior (e.g., oscillations, instability, and divergence) when the problem is multiscale (i.e., when one term in Eq. 9 dominates over the others in some part of $\Omega$; this leads to the occurrence of boundary layers, shocks, chemical fronts, etc.). The present study approaches the solution of Eq. 22 using the Variational Multiscale Finite Element Method (VMS-FEM) [27, 28]. Initial work in the application of the VMS-FEM for LTE and NLTE thermal plasma flows is reported in [13, 14]. The method consists in dividing the solution field $\mathbf{Y}$ into its large-scale $\bar{\mathbf{Y}}$ and small-scale $\mathbf{Y}'$ components, i.e. $\mathbf{Y} = \bar{\mathbf{Y}} + \mathbf{Y}'$, where the large-scales are solved by the computational discretization (computational mesh) and the small-scales, which cannot be described by the discretization, are modeled. Applying the scale decomposition to Eq. 22 leads to [28]:

$$\underbrace{(\bar{\mathbf{W}},\mathcal{R}(\bar{\mathbf{Y}}))_\Omega}_{\text{large scales}} + \underbrace{(-\mathcal{L}^*\bar{\mathbf{W}},\mathbf{Y}')_\Omega}_{\text{small scales}} = \mathbf{0}, \tag{23}$$

where $\mathcal{L} = \mathbf{A_0}\partial_t + (\mathbf{A}_i\partial_i) - \partial_i(\mathbf{K}_{ij}\partial_j) - \mathbf{S_1}$ is the TADR transport operator, and $^*$ the adjoint operator. The small scales are modeled using the residual-based approximation:

$$\mathbf{Y}' = -\boldsymbol{\tau}\mathcal{R}(\mathbf{Y}), \tag{24}$$

where

$$\boldsymbol{\tau} \approx \mathcal{L}^{-1} = (\mathbf{A_0}\partial_t + (\mathbf{A}_i\partial_i) - \partial_i(\mathbf{K}_{ij}\partial_j) - \mathbf{S_1})^{-1} \tag{25}$$

is an algebraic approximation of the integral operator $\mathcal{L}^{-1}$.

Using Eqs. 25 and 24 in Eq. 23, and explicitly separating the temporal discretization from the spatial discretization, the following discrete counterpart to Eq. 23 is obtained:

$$\begin{aligned}
&\mathbf{R}(\mathbf{Y}_h,\dot{\mathbf{Y}}_h) = \\
&\underbrace{(\mathbf{N},\mathbf{A_0}\dot{\mathbf{Y}}_h + \mathbf{A}_i\partial_i\mathbf{Y}_h - \mathbf{S_1}\mathbf{Y}_h - \mathbf{S_0})_{\Omega_h} + (\partial_i\mathbf{N},\mathbf{K}_{ij}\partial_j\mathbf{Y}_h)_{\Omega_h} - (\mathbf{N},n_i\mathbf{K}_{ij}\partial_j\mathbf{Y}_h)_{\Gamma_h}}_{\text{large scales}} + \\
&\underbrace{(\mathbf{A}_i^\mathsf{T}\partial_i\mathbf{N} + \mathbf{S_1}^\mathsf{T}\mathbf{N},\boldsymbol{\tau}(\mathbf{A_0}\dot{\mathbf{Y}}_h + \mathbf{A}_i\partial_i\mathbf{Y}_h - \mathbf{S_1}\mathbf{Y}_h - \mathbf{S_0}))_{\Omega'_h}}_{\text{small scales}} + \\
&\underbrace{(\partial_i\mathbf{N},\mathbf{K}_{ij}^{DC}\partial_j\mathbf{Y}_h)_{\Omega_h}}_{\text{discontinuity capturing}} = \mathbf{0}
\end{aligned} \tag{26}$$

where **R** is the discrete counterpart to $\mathcal{R}$, $\mathbf{Y}_h$ is the discrete representation of **Y**, and $\dot{\mathbf{Y}}_h$ its temporal derivative, **N** is the Finite Element basis function (e.g., see [30]), $\Omega_h$ and $\Gamma_h$ represent the discrete spatial domain and its boundary, respectively, **n** is the outer normal to the boundary, and $\Omega'_h$ is a subset of $\Omega_h$ adequate for the description of the small-scales. The third term in the large-scales component represents the imposition of boundary conditions over $\Gamma$. The discontinuity capturing term has been added to increase the robustness of the solution process in regions with large gradients [38]. Detailed expressions for $\boldsymbol{\tau}$ and $\mathbf{K}^{DC}$ used in the study are presented in [35]. The formulation given by Eq. 26 is second-order accurate in space for linear or multi-linear basis functions (e.g., tetrahedral or hexahedra finite elements) [30] and could be implemented within a standard Finite Elements code, such as Comsol Multiphysics [45].

*3.2. Solution approach*

To obtain second-order accuracy of the overall formulation, the differential-algebraic system given by Eq. 26 is solved using the second-order-accurate generalized-alpha method [31]. Denoting as *n* the time interval of the current solution, the solution at the next time interval *n* + 1 consists in the simultaneous solution of the following system of equations:

$$\mathbf{R}(\mathbf{Y}_{n+\alpha_f}, \dot{\mathbf{Y}}_{n+\alpha_m}) = \mathbf{0}, \tag{27}$$

$$\mathbf{Y}_{n+\alpha_f} = \alpha_f \mathbf{Y}_{n+1} + (1-\alpha_f)\mathbf{Y}_n, \tag{28}$$

$$\dot{\mathbf{Y}}_{n+\alpha_m} = \alpha_m \dot{\mathbf{Y}}_{n+1} + (1-\alpha_m)\dot{\mathbf{Y}}_n, \tag{29}$$

$$\frac{\mathbf{Y}_{n+1} + \mathbf{Y}_n}{\Delta t} = \alpha_g \dot{\mathbf{Y}}_{n+1} + (1-\alpha_g)\dot{\mathbf{Y}}_n, \tag{30}$$

where $\Delta t$ represents the time step size, and $\alpha_f$, $\alpha_m$, and $\alpha_g$ are parameters function of the single user-specified parameter $\alpha \in [0,1]$, and the subscript *h* has been removed from the vectors **Y** and $\dot{\mathbf{Y}}$ to simplify the notation.

Equation 27 implies the solution of a nonlinear system for **Y**, which is solved by an inexact Newton method with line-search globalization given by:

$$\|\mathbf{R}^k + \mathbf{J}^k \Delta \mathbf{Y}^k\| \leq \eta^k \|\mathbf{R}^k\|, \text{ and} \tag{31}$$

$$\mathbf{Y}^{k+1} = \mathbf{Y}^k + \lambda^k \Delta \mathbf{Y}^k, \tag{32}$$

where the super-index *k* represents the iteration counter, $\mathbf{J} \approx \partial \mathbf{R}/\partial \mathbf{Y}$ is the approximate Jacobian, and $\eta$ and $\lambda$ are the tolerance for the solution of the linear system implied by Eq. 31 and the step length, respectively [32]. The solution of Eq. 31 is accomplished using the Generalized Minimal Residual method using the block-diagonal section of **J** as preconditioner [33, 34].

## 4. Simulation set up

*4.1. Spatial domain and discretization*

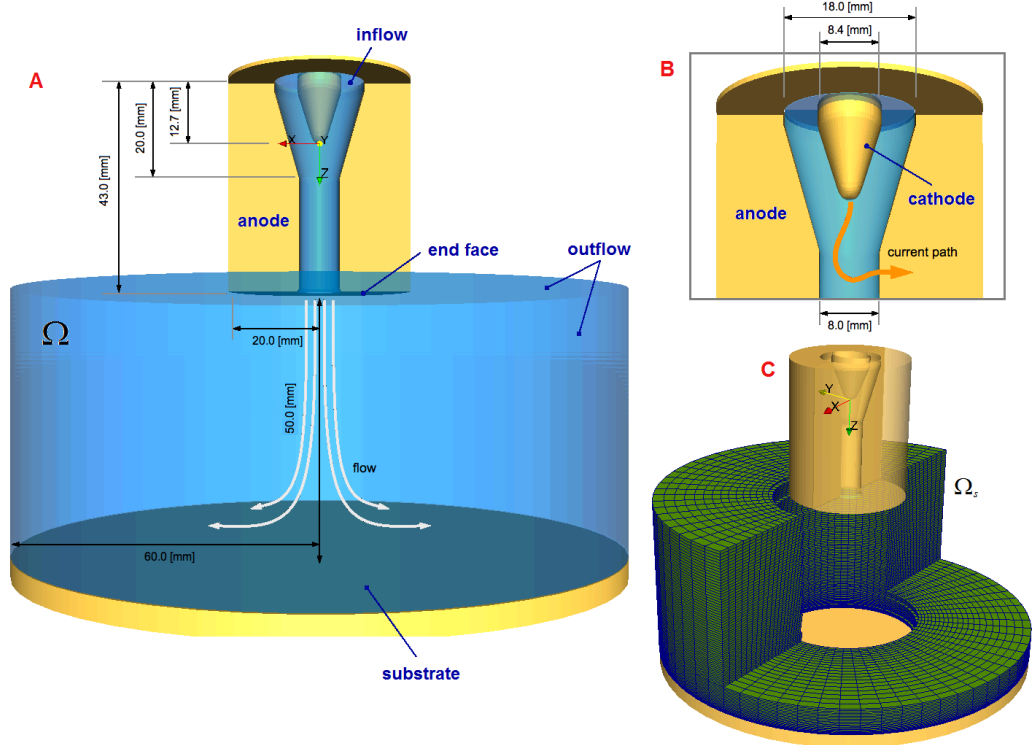

**Figure 2.** Computational domain for the flow from an arc plasma torch: (A) Spatial domain $\Omega$, boundary surfaces (*cathode*, *anode*, *inflow*, *outflow*, *substrate*, and *end face*) and characteristic dimensions; (B) detail of the cathode region and depiction of the current path; and (C) numerical domain of the sponge zone $\Omega_s$ used to mitigate wave reflection from the *outflow* boundary.

Figure 2 depicts the spatial domain $\Omega$ together with its characteristic dimensions, the domain boundary $\Gamma$, and part of the spatial discretization used. The geometry of the torch corresponds to that of a commercial arc plasma torch SG-100 from Praxair Surface Technology, Concord, NH. The inside diameter at the torch outlet is 8.0 [mm] and the distance between the torch and the substrate is 50.0 [mm]. The small distance between the torch and the substrate was chosen to observe more clearly the flow dynamics near the substrate. The computational domain in the radial direction along the substrate extends to 60.0 [mm]. Frame B in Fig. 2 schematically depicts the current path, and therefore the shape of the electric arc. Frame C shows a set of elements used to model the so-called *sponge zone* (domain denoted with $\Omega_s$) used to mitigate wave reflection from the *outflow* boundary (see Section 4.3).

The domain is discretized using a mesh with ~ 370 k nodes and ~ 360 k unstructured hexahedral elements. Due to the fully-coupled numerical model, simulation of the plasma flow requires the coupled solution of ~ 3.7 M nonlinear equations at each time step.

*4.2 Boundary conditions*

The numerical simulations describe the flow of plasma from an inflow of argon entering the torch and discharging into an argon environment. The set of boundary conditions used, consistent with the NLTE model, is listed in Table 2.

**Table 2.** Set of boundary conditions for the flow from a DC arc plasma torch.

| Boundary | Variable | | | | | |
|---|---|---|---|---|---|---|
| | $p$ | $\mathbf{u}$ | $T_h$ | $T_e$ | $\phi_p$ | $\mathbf{A}$ |
| Cathode | $\partial_n p = 0$ | $\mathbf{u} = \mathbf{0}$ | $T_h = T_{cath}(z)$ | $\partial_n T_e = 0$ | $-\sigma \partial_n \phi_p = J_{qcath}(r)$ | $\partial_n \mathbf{A} = \mathbf{0}$ |
| Anode | $\partial_n p = 0$ | $\mathbf{u} = \mathbf{0}$ | $-\kappa_h \partial_n T_h = h_w(T_h - T_w)$ | $\partial_n T_e = 0$ | $\phi_p = 0$ | $\partial_n \mathbf{A} = \mathbf{0}$ |
| Inflow | $\partial_n p = 0$ | $\mathbf{u} = \mathbf{u}_{in}$ | $T_h = T_{in}$ | $T_e = T_{in}$ | $\partial_n \phi_p = 0$ | $\mathbf{A} = \mathbf{0}$ |
| Outflow | $p = p_\infty$ | $\partial_n \mathbf{u} = \mathbf{0}$ | $T_h = T_\infty$ | $T_e = T_\infty$ | $\partial_n \phi_p = 0$ | $\partial_n \mathbf{A} = \mathbf{0}$ |
| Substrate | $\partial_n p = 0$ | $\mathbf{u} = \mathbf{0}$ | $-\kappa_h \partial_n T_h = h_w(T_h - T_w)$ | $\partial_n T_e = 0$ | $\partial_n \phi_p = 0$ | $\partial_n \mathbf{A} = \mathbf{0}$ |
| End Face | $\partial_n p = 0$ | $\mathbf{u} = \mathbf{0}$ | $-\kappa_h \partial_n T_h = h_w(T_h - T_w)$ | $\partial_n T_e = 0$ | $\partial_n \phi_p = 0$ | $\partial_n \mathbf{A} = \mathbf{0}$ |

In Table 2, $\partial_n \equiv \mathbf{n} \cdot \nabla$, with $\mathbf{n}$ as the outer normal, denotes the derivative normal to the surface; $p_\infty$ is the reference open flow pressure, set equal to the atmospheric pressure (1.01 $10^5$ [Pa]), and $\mathbf{u}_{in}$ is the inlet velocity profile given by:

$$\frac{\mathbf{u}_{in}}{U_m} =$$
$$(-\cos\theta_{in}\sin\alpha_{in} - \sin\theta_{in}\cos\alpha_{in}\sin\omega_{in})\hat{\mathbf{x}} +$$
$$(-\sin\theta_{in}\sin\alpha_{in} + \cos\theta_{in}\cos\alpha_{in}\sin\omega_{in})\hat{\mathbf{y}} + \quad (33)$$
$$(\cos\alpha_{in}\cos\omega_{in})\hat{\mathbf{z}}$$

where $U_m$ is the maximum velocity, $\theta_{in} = \theta_{in}(x,y)$ is the polar angle (within the x-y plane), $\alpha_{in}$ is the azimuthal angle (with respect to the z axis), $\omega_{in}$ is the swirl angle (angle with the x-y plane), and $\hat{\mathbf{x}}$, $\hat{\mathbf{y}}$, and $\hat{\mathbf{z}}$ are the unit vectors for each axis. The value of $U_m$ is chosen such that, given total volumetric flow rate $Q_{tot}$, $Q_{tot} = \int_{S_{inlet}} \mathbf{u}_{in} dS$, where $S_{inlet}$ represents the inlet surface. For the simulations presented in Section 5, $Q_{tot} = 10^{-3}$ [m$^3$-s$^{-1}$] (approximately 60 [slpm] of argon) and straight injection ($\alpha_{in} = 0$) with no swirl ($\omega_{in} = 0$) is used.

The temperature profile imposed over the cathode surface $T_{cath}$ is given by:

$$T_{cath} = T_{crod} + (T_{ctip} - T_{crod})\exp(-(z/L_{cath})^2), \quad (34)$$

where $T_{crod}$ and $T_{ctip}$ are the temperatures of the cathode rod and tip, equal to 500 [K] and 3600 [K] respectively, and $L_{cath}$ is a characteristic length set equal to 1.5 [mm]. Heat transfer to solid surfaces is modeled assuming convective heat losses in a water-cooled metal using $h_w = 10^5$ [W-m$^{-2}$-K$^{-1}$] as the convective heat transfer coefficient and $T_w = 500$ [K] as the reference cooling water temperature [11, 19]. $T_\infty = 500$ [K] is a reference open flow temperature and $T_{in} = 500$ [K] is the inflow temperature.

The current density profile over the cathode $J_{qcath}$ is given by:

$$J_{qcath} = J_{qmax}\exp(-(r/r_{cath})^{n_{cath}}), \quad (35)$$

where $r = (x^2 + y^2)^{\frac{1}{2}}$ is the radial coordinate, and $J_{qmax}$, $r_{cath}$, and $n_{cath}$ are parameters that control the shape of the current density profile, which has to satisfy the imposition of the total electric current to the system, i.e. $I_{tot} = \int_{S_{cath}} J_{qcath} dS$, where $S_{cath}$ represents the cathode surface. The simulations presented in Section 5 correspond to a value of $I_{tot} = 400$ [A], using $n_{cath} = 4$, $J_{qmax} = 2.0 \; 10^8$ [A-m$^{-2}$] and $r_{cath} = 0.80918$ [mm].

*4.3. Nonreflecting boundary conditions*

The numerical simulation of fluid flows often requires the truncation of the physical domain (e.g., reaction chamber, surrounding environment) to reduce the computational cost of the simulation. A suitable numerical model has to ensure that the imposition of boundary conditions over the artificial boundary does not appreciably affect the dynamics of the flow under study. This is often accomplished by the use of so-called *artificial boundary conditions* (also known as absorbing, nonreflecting, radiation, invisible, or far-field). In the case of compressible flows, outflow conditions need to ensure the uninterrupted transit of the flow characteristics out of the domain [46]. Typical effects of the use of inadequate outflow conditions are unphysical heating, pressure build-up, and wave reflection. The imposition of zero gradient of the transported variable, probably the simplest and most frequently used outflow condition in plasma flow modeling, is too reflective, especially when the flow approaching the boundary varies significantly in time and space.

Among the many approaches for nonreflecting outflow boundary conditions (e.g., non-linear characteristics, grid stretching, fringe methods, windowing) the use of absorbing layers (*sponge zones*) is one of the simplest and most effective [46]. Given the conservation equation of a variable $\psi$, $\mathcal{R}(\psi) = \partial_t \psi + \nabla \cdot \mathbf{f}_\psi - s_\psi = 0$, where $\mathbf{f}_\psi$ represents the total flux of $\psi$ and $s_\psi$ a source term, specification of a nonreflecting boundary condition requires the modification of $\mathcal{R}(\psi)$ through the *sponge zone* $\Omega_s$ (i.e., part of the spatial domain $\Omega$ near the boundary) according to:

$$\mathcal{R}(\psi) = -\sigma_\psi (\psi - \psi_\infty) , \qquad (36)$$

where $\sigma_\psi = \sigma_\psi(x_n)$ is a damping coefficient that varies spatially in the direction normal to the boundary, $x_n = \mathbf{X} \cdot \mathbf{n}$ is the distance perpendicular to the boundary, and $\psi_\infty$ represents a reference value of $\psi$ (e.g. the value of $\psi$ far from the boundary). The design of $\sigma_\psi$ should ensure a smooth transition from zero in the flow domain to a positive value at the boundary. The region in which $\sigma_\psi$ is greater than 0 is known as the *absorbing layer*. Clearly, if $\sigma_\psi = 0$, then the original conservation equation is recovered and no mitigation of waves reflection is accomplished. A large enough value of $\sigma_\psi$ causes disturbances to decay exponentially and at the same time makes the variable $\psi$ approach $\psi_\infty$. The use of a sponge zone is usually empirical and does not completely prevent wave reflection, but it does allow attenuation of outgoing waves as they travel through the absorbing layer.

For the plasma flow model in Eq. 9, the use of a sponge zone implies that the TADR system is modified as:

$$\mathcal{R}(\mathbf{Y}) = -\boldsymbol{\sigma}_\mathbf{Y} (\mathbf{Y} - \mathbf{Y}_\infty) , \qquad (37)$$

in the region $\Omega_s \supset \Omega$, where $\boldsymbol{\sigma}_\mathbf{Y}$ is the damping matrix, and $\mathbf{Y}_\infty$ is the reference value of the vector of unknowns $\mathbf{Y}$. Figure 2 depicts part of the sponge zone $\Omega_s$ used to mitigate wave reflection as the flow exits the *outflow* boundary. The modification of the set of TADR equations implied by Eq. 37, in general, invalidates the solution of **Y** obtained through $\Omega_s$ (e.g., invalidates total mass conservation); therefore, the solution of **Y** is valid only in the part of the domain $\Omega \setminus \Omega_s$ (i.e., the domain with the sponge zone excluded).

In the present study, the following definitions of $\boldsymbol{\sigma}_Y$ and $\mathbf{Y}_\infty$ were used:

$$\boldsymbol{\sigma}_Y = diag([\ 1\ \ 0\ \ 0\ \ 1\ \ 1\ \ 0\ \ 0\ \ 0\ \ 0\ \ 0\ ]^T)\text{ and} \tag{38}$$

$$\mathbf{Y}_\infty = [\ p_\infty\ \ 0\ \ 0\ \ 0\ \ T_\infty\ \ T_\infty\ \ 0\ \ 0\ \ 0\ \ 0\ ]^T, \tag{39}$$

where *diag* represents the operator such that, for a matrix $\mathbf{A} = [a_{ij}]$ and vector $\mathbf{B} = [b_i]$, $\mathbf{A} = diag(\mathbf{B})$ implies: $a_{ij} = b_i \delta_{ij}$. Equation 38 indicates that the damping through the sponge zone only affects the variables $p$, $u_z$, and $T_h$, whereas Eq. 39 shows that those variables approach the values $p_\infty$, 0, and $T_\infty$, respectively. The form of $\boldsymbol{\sigma}_Y$ given by Eq. 38 was found to be effective to mitigate boundary effects on the simulations presented in Section 5 (e.g., larger values of $\boldsymbol{\sigma}_Y$ caused excessive damping of the solution, whereas smaller values caused pressure build-up and heating around the outflow boundary).

## 5. Flow dynamics from a dc arc plasma jet

*5.1. Arc reattachment and flow dynamics*

The dynamics of the arc inside the torch, as described in Section 1, play a primary role in the flow of the plasma jet. The arc dynamics are a consequence of the unstable imbalance between the Lorentz force exerted over the arc due to the distribution of current density connecting the anode and cathode attachments, and the drag caused by the relatively cold and dense stream of inflow working gas over the hot and low-density arc plasma. The force imbalance causes the dragging of the anode attachment, the elongation of the arc, and the increase of the arc curvature until the arc gets in close proximity to another location over the anode. If the arc gets in contact with a location over the anode closer to the cathode, the arc will tend to attach, forming another anode attachment. The new attachment, being more energetically favorable than the old one (i.e., if the arc cross section remains constant, a smaller voltage drop is needed to transfer the same amount of current), prevails and the old attachment eventually fades away. This process is customarily known as the *arc reattachment process* (e.g., [11, 19, 48]), and is responsible of the quasi-periodic voltage signal from non-transferred arc plasma torches [3, 4].

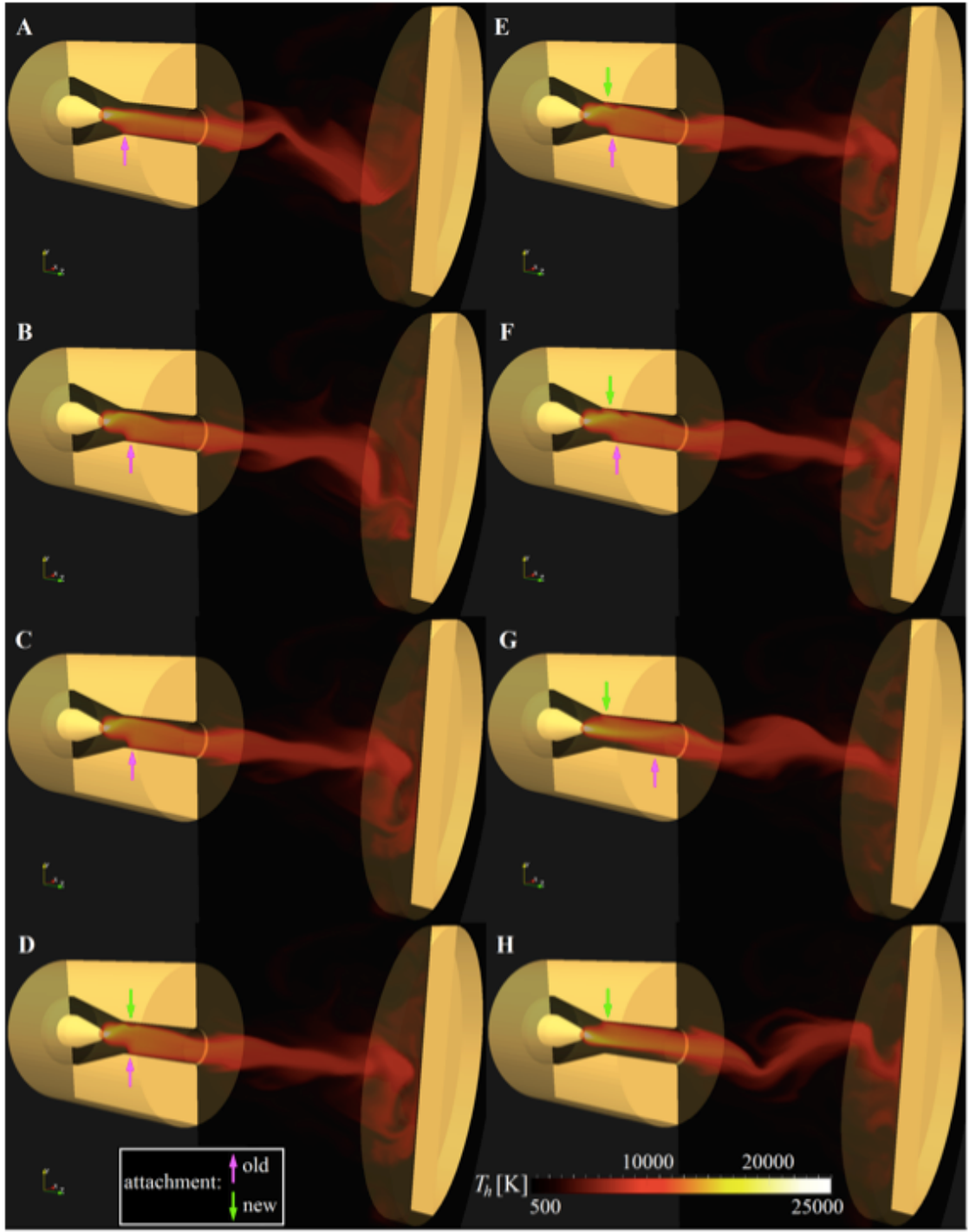

**Fig. 3.** Flow dynamics during an arc reattachment event: Sequence of snapshots of $T_h$ distribution. The arrows indicate the location of the initial (*old*) and formed (*new*) arc anode attachments. The total voltage drop $|\phi_{p,\max} - \phi_{p,\min}|$ is: (A) 32.6, (B) 34.2, (C) 35.6, (D) 32.5, (E) 31.5, (F) 29.8, (G) 31.2, and (H) 32.3 [V]. (See animation 1 in supplemental materials.)

The behavior described above can be observed in the results in Fig. 3, which presents a sequence of snapshots of the distribution of heavy-species temperature $T_h$ over the domain during an *arc reattachment* event. Only the part of the domain that does not include the sponge zone is shown (recall from Section 4.1 that the spatial domain extends 60 [mm] radially along the substrate). In Fig. 3, the locations of the original (*old*) and newly formed (*new*) attachments are indicated with arrows. Due to the use of straight injection, the new attachment forms approximately at the opposite end of the old attachment along the same plane that crosses the torch axially. Frames A to C in Fig. 3 show the progressive movement of the *old* anode attachment and the corresponding increase of the arc length and curvature. Figure 3 – frame D indicates the first instance in which the two attachments coexist, and frames E to G show how the *new* attachment gets established whereas the *old* one fades away. Figure 3 – frame H indicates that the *new* attachment remains, starts being dragged by the gas, which will eventually lead to the start a new reattachment cycle. The flow dynamics during the repeated reattachment events are presented animation 1 in the supplemental materials.

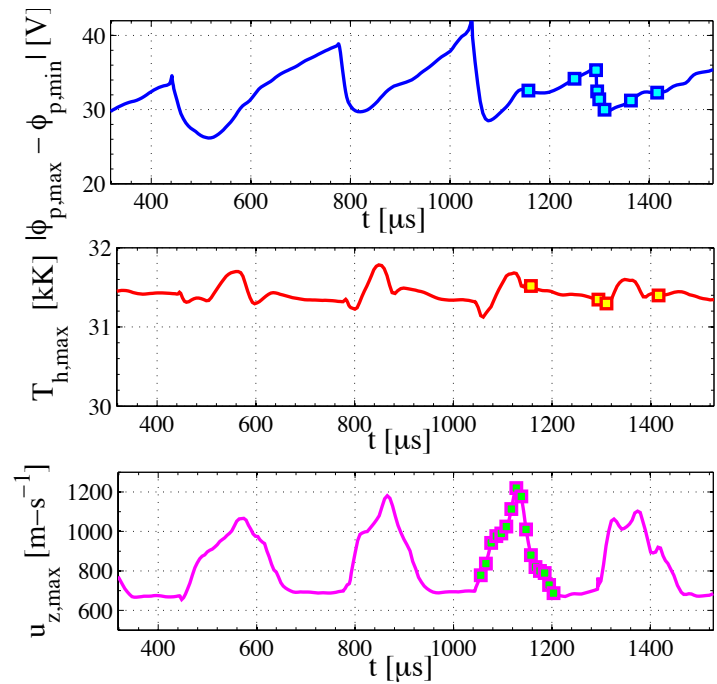

**Fig. 4.** Temporal signals: (*top*) total voltage drop, (*center*) maximum heavy-species temperature, and (*bottom*) maximum axial velocity. The symbols indicate the locations of the snapshots presented in: (*top*) Fig. 3; (*center*) Figs. 5, 6, and 8; and (*bottom*) Fig. 7.

The effect of the formation of the new attachment in also evidenced by the evolution of the voltage drop across the electrodes $\Delta\phi_p = |\phi_{p,max} - \phi_{p,min}|$, which shows that the voltage drop is maximum right before the new attachment forms (e.g., the location of the old anode attachment is furthest away from the cathode) and that the formation of the new attachment causes a significant drop in voltage (i.e., ~ 15% decrease in $\Delta\phi_p$ between frames C and F). The evolution of $\Delta\phi_p$, together with the evolution of the maximum heavy-species temperature and maximum axial velocity, are presented in Fig. 4 - *top*. The locations corresponding to the snapshots in Fig. 3 are indicated by symbols in the plot of $|\phi_{p,max} - \phi_{p,min}|$, which also evidence the quasi-periodic nature of the arc dynamics. The magnitude and periodicity of the voltage signal are consistent with the experimental results reported in [4].

*5.2. Thermodynamic nonequilibrium*

Figure 5 depicts the differences between the heavy-species and electron temperature fields during an arc reattachment event. The time instants corresponding to the snapshots in Fig. 5 are indicated by symbols in the plot of $T_{h,max}$ in Fig. 4 – *center*. The results in Fig. 4 – *center* indicate that the maximum temperature remains relatively constant (~ 31.5 [kK]) and displays a relative minimum after each arc reattachment event. The temperatures along the jet are consistent with the experimental values reported in [49] (i.e., ~ 12500 K for a torch operating with 400 A).

Similarly to Fig. 3, the arrows in Fig. 5 approximately indicate the location of the anode attachments. Whereas the distribution of both, $T_h$ and $T_e$, show the large-scale undulating characteristics of the jet, the distribution of $T_h$ displays some fine scales features not observed in the $T_e$ distribution. The distribution of $T_e$ is clearly more diffusive, which is in part explained by the higher thermal conductivity of electrons than that for the heavy-species [41, 42]. The results in Fig. 5 also show that the electron temperature remains relatively high far from the jet. The degree of thermodynamic nonequilibrium, i.e. the deviation between $T_h$ and $T_e$, is addressed in the next section.

The comparison between the location of the anode attachment and the curvature of the jet in Fig. 5 - frame 1 with those in Fig. 5 - frame 4, for both temperature fields, appears to evidence a correlation between the dynamics of the arc inside the torch and the large-scale structure of the jet. Such correlation is subtle and can be explored in greater detail by observing the results in animation 1 in supplemental materials.

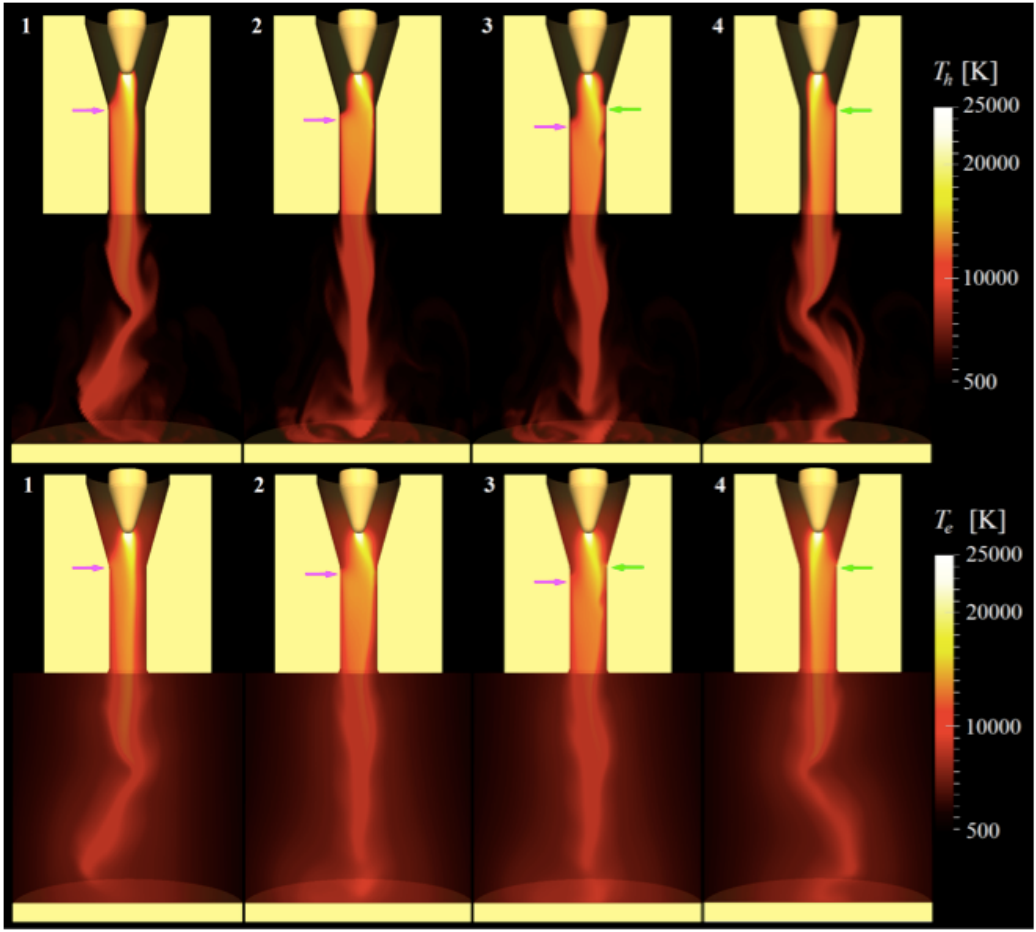

**Fig. 5.** Temperature dynamics during an arc reattachment event: Sequence of (*top*) $T_h$ and (*bottom*) $T_e$ distributions. The $T_e$ field is markedly more diffuse than the $T_h$ field and does not evidence the development of fluid dynamic instabilities. The arrows indicate the location of arc anode attachments.

*5.3. Thermodynamic, electrical, and fluid relaxation processes*

A plasma flow experiences different types of relaxation processes that drive it towards a state of equilibrium (i.e., zero fluxes) in the regions away from the forcing sources (e.g., stream of working gas, imposed total current). Three of these relaxation processes can be assessed by the results in Fig. 5, namely: (1) the relaxation of electron and heavy-species energy, which is modeled by the $K_{eh}(T_e - T_h)$ term in Table 1; (2) electrical relaxation manifested by the decay of current density $\mathbf{J}_q$ far from the electrodes due to the nature of the charge conservation equation $\nabla \cdot \mathbf{J}_q = 0$ and the imposed boundary conditions for $\phi_p$ in Table 2; and (3) the dissipation of linear momentum $\rho\mathbf{u}$ caused by fluid viscosity $\mu$. The temporal locations of the snapshots are the same used in Fig. 5 and indicated by the symbols in the plot of $T_{h,max}$ in Fig. 4 - *center*.

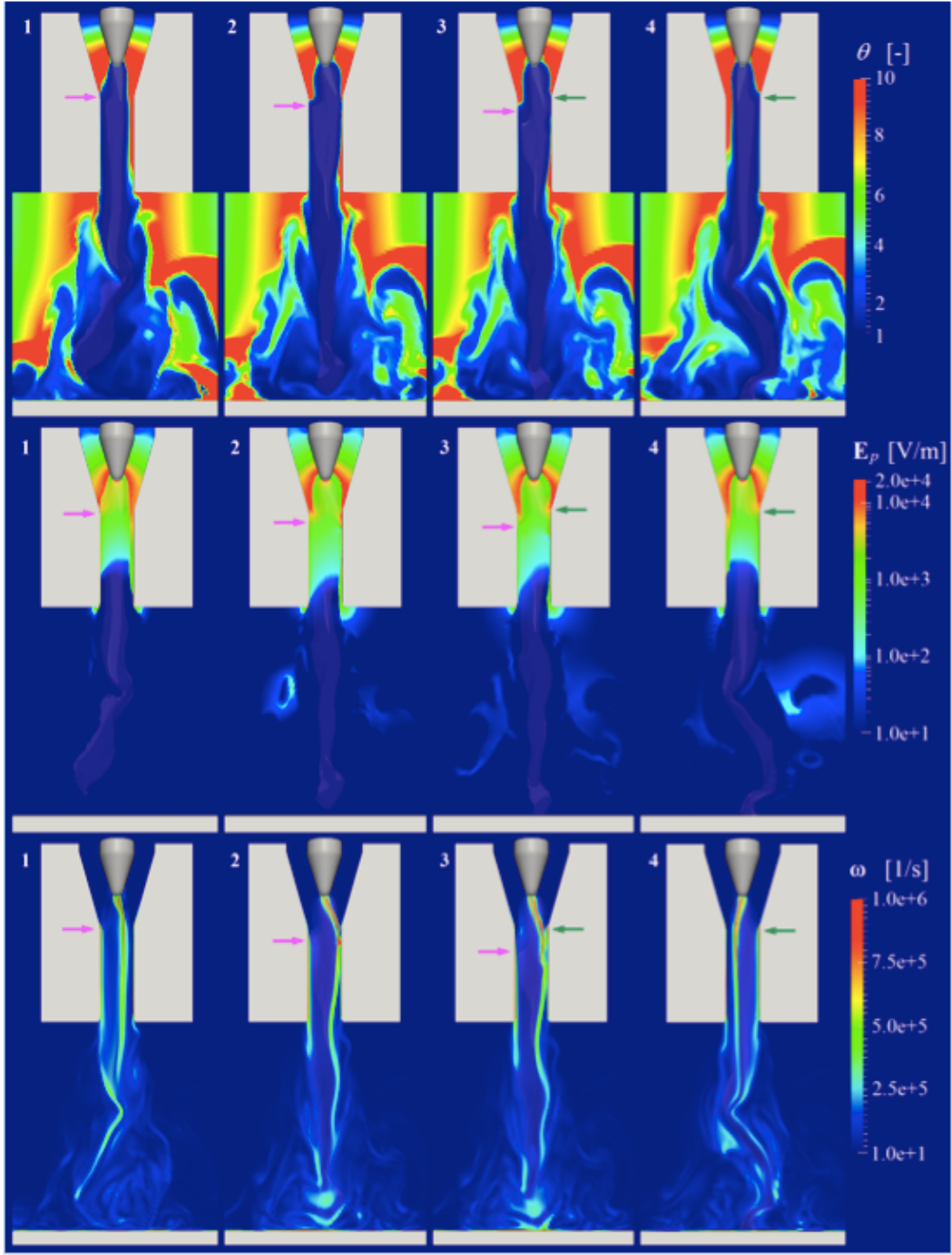

**Fig. 6.** (*Top*) thermodynamic nonequilibrium parameter $\theta = T_e/T_h$, (*center*) magnitude of effective electric field $||\mathbf{E}_p||$, and (*bottom*) magnitude of vorticity $||\boldsymbol{\omega}||=||\nabla \times \mathbf{u}||$ during an arc reattachment event. The arrows indicate the location of arc anode attachments.

The electron - heavy-species energy relaxation is depicted by the distribution of the thermodynamic nonequilibrium parameter $\theta = T_e/T_h$ in Fig. 6 - *top*. The core of the plasma shows $\theta \sim 1$ (i.e., thermodynamic equilibrium), whereas the maximum values of $\theta$ are found very close to the plasma fringes, which subsequently decay to values of $\theta \sim 1$ far from the plasma. The complex distribution of $\theta$ is a consequence of the markedly dissimilar distribution of $T_h$ and $T_e$ around the arc fringes, as depicted in Fig. 5.

The results in Fig. 6 – *center* show the distribution of the effective electric field $\mathbf{E}_p$. A more detailed description of the variation of electric field inside the torch during an arc reattachment process is presented in [14]. The results indicate that, as expected, the magnitude of the electric field $\|\mathbf{E}_p\|$ rapidly decays away from the arc. Interestingly, the results also show that the fluctuation of the jet cause the occurrence of localized pockets with relatively high $\|\mathbf{E}_p\|$ away from the plasma jet (e.g., at the right side of the jet in Fig. 6 – *center* – frame 4).

The images in Fig. 6 – *bottom* show the distribution of vorticity magnitude $\|\boldsymbol{\omega}\| = \|\nabla \times \mathbf{u}\|$. The high values of $\|\boldsymbol{\omega}\|$ along the anode surface are a consequence of the large velocity gradients near the anode surface. The high values of $\|\boldsymbol{\omega}\|$ in the jet fringes could be explained by the high generation of vorticity due to the large mass density gradients $\nabla \rho$ there (vorticity generation can be expressed by the term $-\rho^{-2} \nabla p \times \nabla \rho$; see [47], Chapter 2). The high magnitude of $\|\boldsymbol{\omega}\|$ in the periphery of the plasma jet is also indicative of large shear, which added to the large temperature and material properties gradients, causes the development of shear flow instabilities, as described in the next section.

*5.4. Instabilities development*

The turbulent nature of the jet from a DC non-transferred arc plasma torch could be arguably be originated in the development of shear flow instabilities (also known as Kelvin-Helmholtz) near the torch exit. Figure 7 shows a sequence of snapshots of the distribution of heavy-species temperature near the torch exit depicting the formation and evolution of a shear instability. The arrows in Fig. 7 point to the location of an instability, and the temporal instants corresponding to the snapshots are indicated by the symbols in the plot of $u_{z,max}$ in Fig. 4 - *bottom*. The results in Fig. 7 are complemented by animation 2 in the supplemental materials.

The inherently sensitive nature of instability development phenomena makes imperative the use of high-accuracy numerical simulations (e.g., using discretization schemes with low numerical dissipation and/or very fine meshes) for their description. The relatively coarseness of the instability captured in Fig. 7 seems to indicate that the onset of the instability and its development may vary significantly if higher numerical resolution is used (e.g., using a higher than second-order-accurate discretization or more discretization nodes).

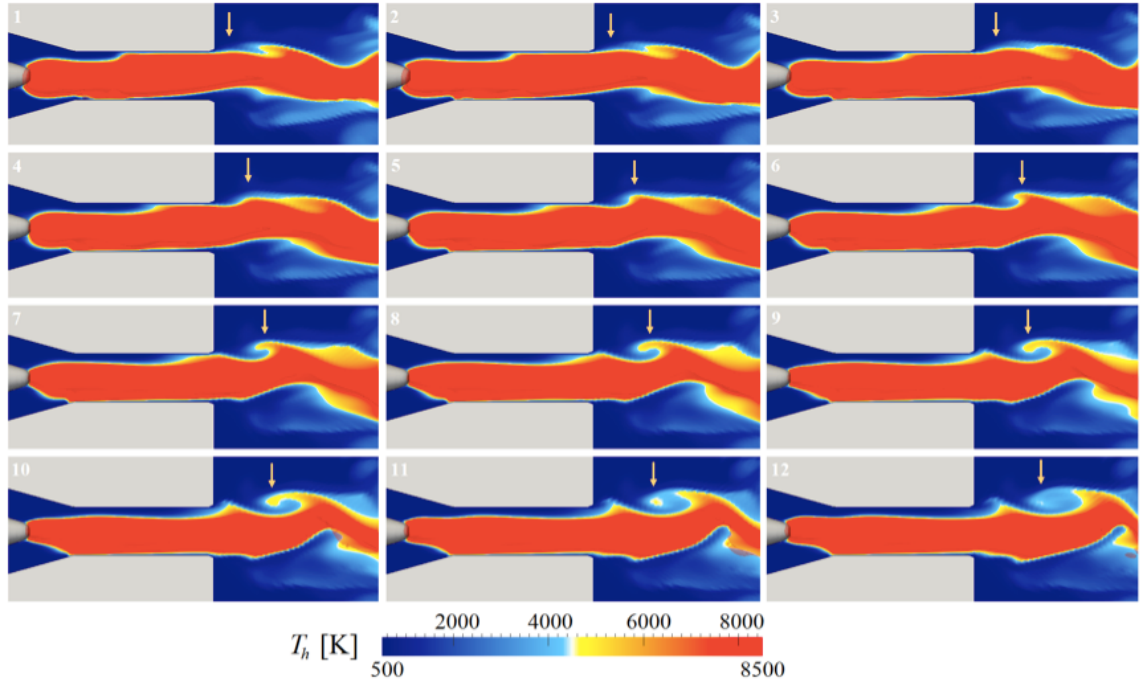

**Fig. 7.** Formation and development of shear flow instabilities: Sequence of snapshots of $T_h$ distribution in the region encompassing the cathode tip and the torch discharge. The arrows indicate the location where an instability develops. The colorscale used emphasizes the development of the instability. (See animation 2 in supplemental materials.)

By contrasting the results in Fig. 5 with those in Fig. 7, it is observed that the shear flow instabilities are correlated with the distribution of $T_h$ and not with the distribution of $T_e$. This finding may indicate, for example, that experimental diagnostics that rely on electron energy (or temperature) only may not be able to capture the onset of shear instabilities near the arc fringes.

*5.5. Fluid dynamics and coherent structures*

Figure 8 presents the evolution of flow streamlines colored with flow speed during a reattachment event. The results indicate a quasi-periodic expansion/contraction of the gas ejected with the plasma jet evidenced by the relative separation of streamlines. This behavior is in part caused by the arc dynamics: the cross section of the plasma is larger near a reattachment event due to the coexistence of two anode attachments, the larger volume of plasma constricts the incoming cold gas causing higher gas velocities (contracted streamlines in Fig. 8 – frames 2 and 3); whereas far from a reattachment event, the incoming gas encounter less opposition from the plasma, and therefore expands more freely (progressive separation of streamlines in Fig. 8 – frames 1 and 4).

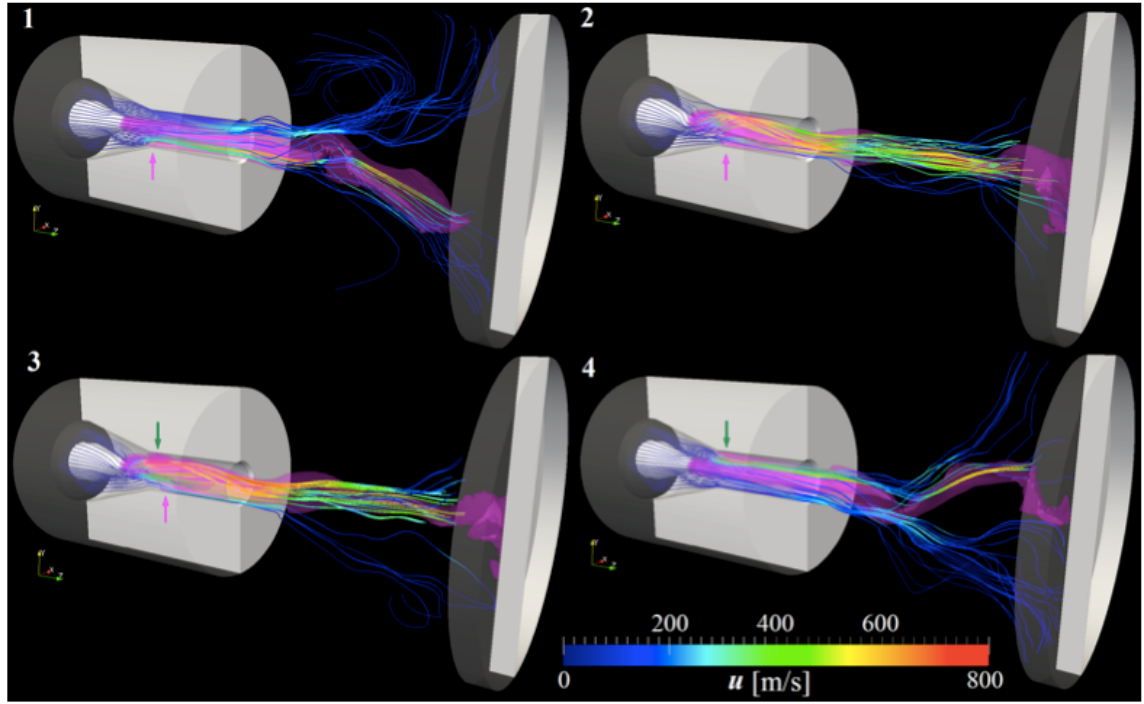

**Fig. 8.** Evolution of flow streamlines during an arc reattachment event. Streamlines are colored with velocity magnitude and superposed over semi-transparent $T_h$ isocontours at 8, 12, and 16 [kK]. The arrows indicate the location of arc anode attachments.

The results in Fig. 8 can be contrasted with the plots of $u_{z,max}$ in Fig. 4 – *bottom*. The maximum velocity $u_{z,max}$ occurs in the so-called *cathode jet*, the region in front of the cathode tip where the plasma gets accelerated due to the electromagnetic pinch caused by the constriction of current density. The variation of $u_{z,max}$ with time shows a positive correlation with the variation of $T_{h,max}$ and a negative correlation with the variation of $\Delta\phi_p$. Therefore, the maximum velocities in the cathode jet are correlated with the expansion of streamlines outside the torch, consistent with the fluctuating behavior of the volume of plasma due to reattachment events.

The complex dynamics from the flow of a DC arc plasma jet often lead to the development of turbulence. The complexity of turbulent and high-vorticity flows has prompted to the development of methods capable to identify so-called *coherent structures* (i.e., large-scale flow features that can be unambiguously be identified) to aid their study. Even though the predominance of turbulence in thermal plasma flows, to the best knowledge of the author, the report in [26] is the only study so far of the identification of coherent flow structures from numerical simulations of a thermal plasma flow.

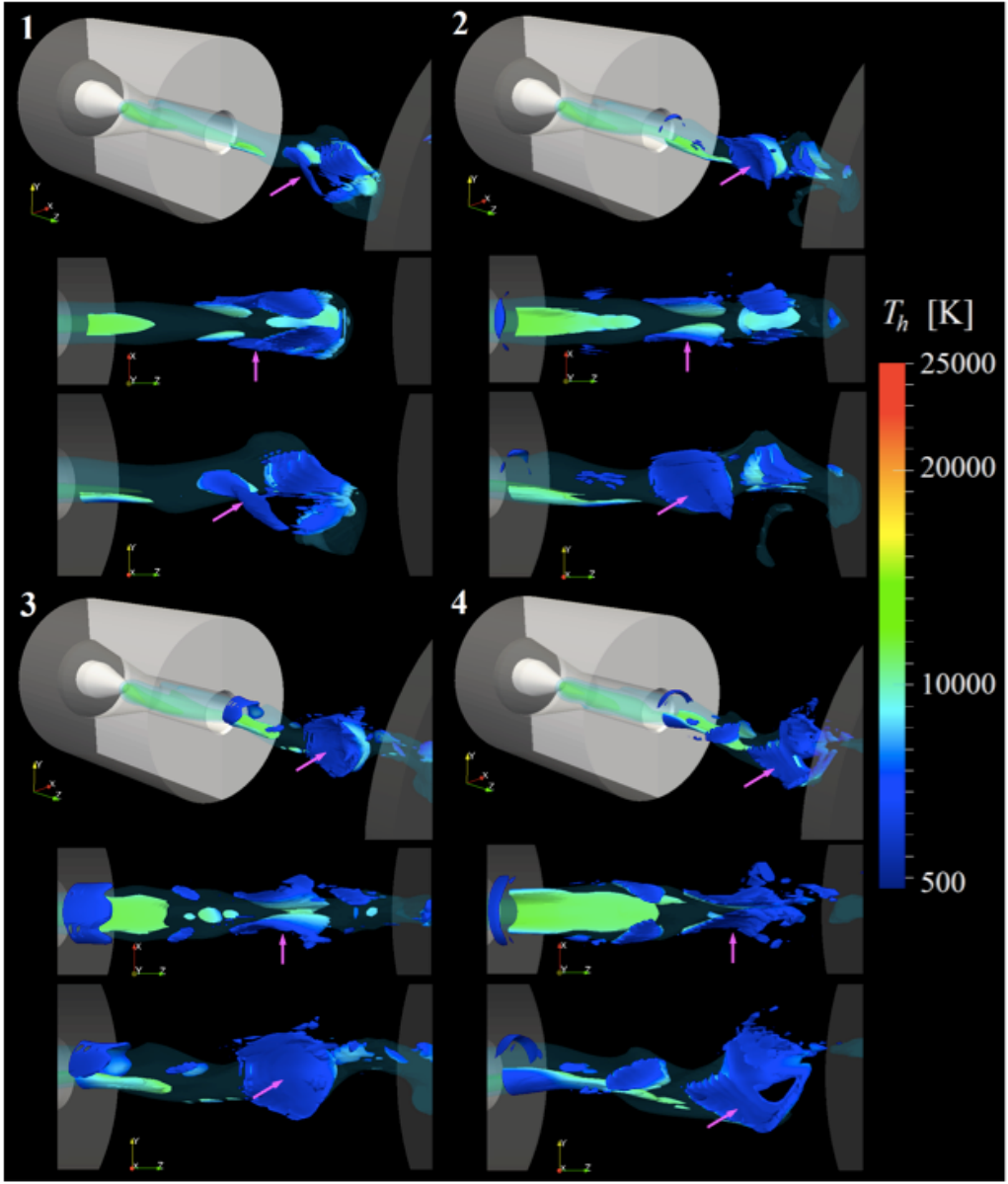

**Fig. 9.** Formation and evolution of coherent structures using the $Q$-criterion for a value of $Q/Q_0 = 2$ ($Q_0$ calculated using $u_{ref} = 500$ [m-s$^{-1}$] and $L_{ref} = 5$ [mm]). Each frame shows three-dimensional, $x$-$z$ and $y$-$z$ views of the structures, and the arrows point to a single lobular structure. The structures are colored with the $T_h$ field. (See animation 3 in supplemental materials.)

The *Q*-criterion for structure identification, used in [26], is derived from the analysis of the distribution of the quantity *Q* defined by [50-52]:

$$\frac{Q}{Q_0} = \tfrac{1}{2}((\mathbf{\Omega}:\mathbf{\Omega}) - (\mathbf{S}:\mathbf{S})), \quad (40)$$

where $\mathbf{\Omega} = \tfrac{1}{2}(\nabla \mathbf{u} - \nabla \mathbf{u}^T)$ and $S = \tfrac{1}{2}(\nabla \mathbf{u} + \nabla \mathbf{u}^T)$ are the symmetric and anti-symmetric parts of the tensor $\nabla \mathbf{u}$ (i.e., $\nabla \mathbf{u} = \mathbf{\Omega} + \mathbf{S}$), and $Q_0 = (u_{ref}/L_{ref})^2$, with $u_{ref}$ and $L_{ref}$ as reference velocity and length, respectively, is a reference value used to normalize *Q*. The *Q*-criterion allows the identification of regions where rotation dominates strain in the flow [51].

Figure 9 shows coherent structures from the plasma jet defined by the isosurface of $Q/Q_0$ = 2, where the reference value $Q_0$ is calculated using $u_{ref}$ = 500 [m-s$^{-1}$] (~ ½ of the maximum velocity in the jet, see Fig. 7 or $u_{z,max}$ in Fig. 4) and $L_{ref}$ = 5 [mm] (recall the inside diameter at the torch exit is 4 [mm], see Fig. 2). The results in Fig. 9 illustrate the formation and evolution of lobular structures at both side of the jet, indicated by arrows through three-dimensional, *z-x* and *z-y* views. A pair of lobular structures forms near the torch exit. The lobes are initially elongated (frame 1); subsequently rotate (frames 1 and 2, *z-y* view), then expand and become more rounded (frames 2 and 3, *z-y* view), and finally break down into smaller structures (frames 3 and 4). The formation and evolution of the structures are more clearly observed in animation 3 in the supplemental materials. It can be observed that the occurrence of coherent structures cannot be deduced from the distribution of vorticity in Fig. 6 - *bottom*.

The structures presented in Fig. 9 appear consistent with the structures determined from experimental data by Hlína and collaborators [6-8], even though the VMS-FEM method used in the present computational study does not constitute a complete LES model and that the limited resolution and accuracy of the simulations prevents them to be considered DNS. Nevertheless, it is expected that the obtained structures and their dynamics may differ considerably in higher resolution simulations.

## 6. Summary and conclusions

The flow dynamics from an arc discharge plasma jet have been studied computationally using time-dependent three-dimensional simulations encompassing the dynamics of the arc inside the torch, the evolution of the jet through the discharge environment, and the later impingement of the jet over a substrate. The plasma is modeled using a thermodynamic nonequilibrium (two-temperature) model together with a Variational Multiscale Finite Element Method. The simulation results show diverse aspects of the flow dynamics, including the forcing of the plasma jet caused by the dynamics of the electric arc inside the torch, the markedly more diffuse distribution of electron temperature with respect to the heavy-species temperature and the prevalence of high degrees of thermodynamic nonequilibrium in the plasma fringes, the development of shear flow instabilities around the jet captured by the distribution of heavy-species temperature, the occurrence of localized regions with high electric fields far from the arc, the fluctuating expansion of the gas ejected from the torch, and the formation and evolution of coherent flow structures. The computational model did not included a comprehensive treatment of flow turbulence, largely found in these types of flows, and therefore it can be expected that simulations with higher accuracy (e.g., finer mesh and/or higher order of accuracy) may capture a more detailed picture of the flow dynamics; particularly of the development of shear instabilities, coherent structures, and the subsequent development of turbulence.